\documentstyle[preprint,prl,aps,epsfig,amssymb,twoside]{revtex}
\begin{document}
\draft
\tighten

\title{Depletion potential in hard sphere fluids}

\author{B. G\"otzelmann, R. Roth, and S. Dietrich}
\address{Fachbereich Physik, Bergische Universit\"at Wuppertal, D-42097
  Wuppertal, Germany}
\author{M. Dijkstra and R. Evans}
\address{H.H. Wills Physics Laboratory, University of Bristol, Bristol BS8 1TL,
United Kingdom}
\maketitle
\begin{abstract}
A versatile new approach for calculating the depletion potential in a hard sphere
mixture is presented. This is valid for any number of components and for arbitrary
densities. We describe two different routes to the depletion potential for the
limit in which the density of one component goes to zero. Both routes can be
implemented within density functional theory and simulation. We illustrate the
approach by calculating the depletion potential for a big hard sphere in a fluid
of small spheres near a planar hard wall. The density functional results are in 
excellent agreement with simulations.
\end{abstract}


\narrowtext

When the separation between two big colloidal particles suspended in a fluid of 
small colloidal particles or non-adsorbing polymers is less than the diameter of 
the small ones exclusion or depletion of the latter occurs leading to anisotropy of
the local pressure which gives rise to an attractive depletion force between the
big particles. Asakura and Oosawa \cite{Asakura54} first described this depletion
mechanism, suggesting that it would drive phase separation in colloid-polymer
mixtures. Using excluded volume arguments they calculated the force for two
hard spheres of radius $R_b$ in a fluid of small hard spheres of radius $R_s$ and
showed that this is attractive for all separations less than $2 R_s$ and is zero 
for larger separations. The hard sphere model captures the essence of the depletion
phenomenon and can be mimicked experimentally by suitable choices of colloidal
solutions \cite{experiments,Dinsmore98}. Depletion forces are of fundamental 
statistical mechanical interest since they are purely entropic in origin 
and much theoretical \cite{Attard89,Attard90,Mao95,Goetzelmann98b}
and simulation \cite{Biben96,Dickman97} effort has been devoted to their
determination. Many recent experiments, using a wide variety of
techniques, have investigated depletion forces for colloidal mixtures or for
colloid-polymer mixtures \cite{experiments}. Much of the impetus for these studies
stems from a need to understand how depletion determines phase separation and
flocculation phenomena and this has stimulated a growing interest in ascertaining
quantitative details of the depletion force. Several of these studies are
concerned with a big particle near a planar wall but Dinsmore et al. 
\cite{Dinsmore98} have employed video microscopy to monitor the position of a 
single big colloid immersed in a solution of small colloids inside a vesicle --
a system resembling hard spheres confined inside a hard cavity. The experimental
conditions in Ref.\cite{Dinsmore98} correspond to a small sphere packing fraction
$\eta_s=0.3$, which is sufficiently high that one expects the depletion force to
be substantially different from that given by the Asakura-Oosawa result; the latter
is valid only in the limit $\eta_s\to 0$. Thus, in order to model this and other
experimental situations one requires a theory of depletion which is reliable
at high packings and which can tackle rather general ``confining'' geometries. The
latter could represent a planar wall, a wedge or cavity 
or, indeed,
another big particle \cite{comment3}. Here we present such a theory for the
depletion {\em potential} based on a density functional treatment (DFT) of a
fluid mixture. Our approach is versatile \cite{comment4} and avoids the
limitations of the virial expansion \cite{Mao95} and of the Derjaguin approximation
\cite{Mao95,Goetzelmann98b} and avoids many of the limitations inherent in
integral-equation theories \cite{Attard89,Attard90,Biben96} of the depletion force.
It has a distinct advantage over a direct implementation of DFT in that it does not 
require a minimization of the free energy in the presence of the big particle. It 
also suggests an alternative simulation procedure for calculating depletion 
potentials. We demonstrate the accuracy of our approach by comparing our results 
for the depletion potential between a big hard sphere and a planar hard wall
with those of our new simulation and with those of Ref.\cite{Dickman97}.

We consider a multicomponent mixture in which species $i$ ($i=1,2,\dots \nu$)
has a chemical potential $\mu_i$ and is subject to an external potential 
$V_i({\bf r})$ giving rise to the equilibrium number density profiles 
$\{\rho_i({\bf r})\}$. The quantity of
interest $W_t({\bf r}_i)\equiv \Omega_{ti}({\bf r}_i;\{\mu_i\},T)-
\Omega_{ti}(\infty;\{\mu_i\},T)$ is the difference in grand potential $\Omega_{ti}$
between a test particle of species $i$ located at position ${\bf r}_i$ and one
located at ${\bf r}_i=\infty$, i.e., deep in the bulk, far from the object
exerting the external potentials. Using the potential distribution theorem in
the grand ensemble one can easily show \cite{Henderson83}:
\begin{equation} \label{PDT}
W_t({\bf r}_i)=-k_B T \ln \left(\frac{\rho_i({\bf r}_i)}{\rho_i(\infty)}\right) - 
V_i({\bf r}_i) +
V_i(\infty),
\end{equation}
where $T$ is the temperature. Equation~(\ref{PDT}) is a general result; it is valid
for arbitrary interparticle forces. It is important for subsequent
application to note that $\rho_i({\bf r})$ is determined solely by the external
potentials so it has the symmetry dictated by these potentials and not the broken
symmetry brought about by inserting the test particle. In order to obtain the
depletion potential $W$, which pertains to a single (big) particle, $i\equiv b$,
in the presence of a fixed object exerting a potential $V_b({\bf r})$ of finite
range, we take the limit $\mu_b\to -\infty$ so that the density of the big particles
vanishes, with the other chemical potentials
$\{\mu_{i\not=b}\}$ fixed. It follows that for ${\bf r}$ outside the range of
$V_b({\bf r})$
\begin{equation} \label{dep}
\beta W({\bf r})=-\lim_{\mu_b\to -\infty} \ln\left(\frac{\rho_b({\bf r})}
{\rho_b(\infty)}\right)
\end{equation}
with $\beta=(k_B T)^{-1}$. Although both $\rho_b({\bf r})$ and $\rho_b(\infty)$
vanish in this limit their ratio is non-zero. Moreover, the density profiles of the
other species $\{\rho_{i\not=b}({\bf r})\}$ reduce to those in the $\nu-1$ component
fluid. Thus, for a binary mixture of big and small ($s$) particles,
$\rho_s({\bf r})$ reduces to the profile of a pure fluid of small particles.
Equation~(\ref{dep}) can be re-stated in a more familiar form as
$p({\bf r})/p(\infty)=\exp(-\beta W({\bf r}))$, where $p({\bf r})$ is the
probability density of finding the big particle at a position ${\bf r}$ from the
fixed object.

There are two distinct ways of implementing this route to the depletion potential.
First one can calculate the density profile $\rho_b({\bf r})$ and, hence,
$W_t({\bf r})$ in the mixture for decreasing concentration of species $b$. For
sufficiently negative values of $\mu_b$, $W_t({\bf r})$ should approach its limiting
value $W({\bf r})$. Such a procedure is straightforward to implement in any
approximate DFT, especially for objects with planar or radial symmetry (e.g.,
planar walls, spherical cavities or a fixed spherical particle). Second one
can attempt to proceed immediately to the limit of a single big particle,
thereby obtaining $W({\bf r})$ directly without evaluation of $\rho_b({\bf r})$.
Equation~(\ref{PDT}) can be re-expressed as a difference in one-body direct
correlation functions \cite{Henderson83}
\begin{equation} \label{PDT2}
-\beta W_t({\bf r}_i)=c_i^{(1)}({\bf r}_i;[\{\rho_i\}])-c_i^{(1)}(\infty;
[\{\rho_i\}])
\end{equation}
and the depletion potential can be expressed formally as
\begin{equation}
-\beta W({\bf r}) = c_b^{(1)}({\bf r};[\{\rho_{i\not=b}\},\rho_b=0])-
 c_b^{(1)}(\infty;[\{\rho_{i\not=b}\},\rho_b=0]).
\end{equation}
In order to implement this result an explicit prescription for $c_b^{(1)}$ is 
required. Within the context of DFT $c_b^{(1)}({\bf r}_i;[\{\rho_i\}])=
-\beta \delta {\cal F}_{ex}[\{\rho_i\}]/\delta \rho_i({\bf r}_i)$, where
${\cal F}_{ex}[\{\rho_i\}]$ is the excess (over the ideal gas) free energy
functional of the inhomogeneous mixture \cite{Evans79}. For certain classes of
(approximate) functionals explicit results can be obtained for $c_i^{(1)}$
which permit the limit $\mu_b\to -\infty$ to be taken \cite{Goetzelmann98a}. An
important example is Rosenfeld's Fundamental Measure functional for hard sphere
mixtures \cite{Rosenfeld89},
\begin{equation}
{\cal F}_{ex}[\{\rho_i\}]=\int d^3 r \Phi(\{n_\alpha({\bf r})\}),
\end{equation}
where the Helmholtz free energy density $\Phi$ is a function of the set of
weighted densities
\begin{equation} \label{weights}
n_\alpha({\bf r})=\sum_{i=1}^{\nu} \int d^3 r' \rho_i({\bf r}') \omega_i^\alpha
({\bf r}-{\bf r}')
\end{equation}
and explicit expressions are available \cite{Rosenfeld89} for the weight functions
$\omega_i^\alpha$ and for $\Phi$. In this theory the free energy of the homogeneous 
mixture is identical to that from Percus-Yevick or scaled-particle theory. Within 
this approach Eq.~(\ref{PDT2}) reduces to
\begin{equation} \label{deppot}
\beta W_t({\bf r})=\int d^3 r' \sum_\alpha \Psi^\alpha({\bf r}') \omega_b^\alpha
({\bf r}'-{\bf r})
\end{equation}
with $i\equiv b$ and
\begin{equation} \label{deriv}
\Psi^\alpha({\bf r}')\equiv \left(\frac{\beta \partial \Phi(\{n_\alpha\})}
{\partial n_\alpha}\right)_{{\bf r}'}-\left(\frac{\beta \partial
\Phi(\{n_\alpha\})} {\partial n_\alpha}\right)_{\infty},
\end{equation}
i.e., the grand potential difference consists of a sum of convolutions of the
functions $\Psi^\alpha$ and the weight functions $\omega_b^\alpha$ of the big
particle. The index $\alpha$ labels $4$ scalar plus $2$ vector weights
\cite{Goetzelmann98a,Rosenfeld89}. Since the derivatives in Eq.~(\ref{deriv}) are
evaluated at equilibrium, the weighted densities must be obtained from 
Eq.~(\ref{weights}), once the density profiles $\rho_i({\bf r})$ are obtained by
minimizing the free energy functional. However, for the binary mixture the
limit $\rho_b({\bf r})\to 0$ implies that $n_\alpha({\bf r})$ involves only 
$\rho_s({\bf r})$ which is given by the solution of the Euler-Lagrange equation
for the pure small-sphere fluid. The depletion potential $W({\bf r})$ is then
given by Eq.~(\ref{deppot}) with $n_\alpha({\bf r})$ determined in this way. Both
schemes have been used to calculate the depletion potential for a big hard sphere,
immersed in a sea of small spheres, near a planar hard wall. Before describing
our results we remark upon two limiting cases of the DFT treatment. 

(i) The present functional \cite{Rosenfeld89} reduces to the exact low density limit
$\beta {\cal F}_{ex}[\{\rho_i\}]=-\frac{1}{2}\sum_{i,j}\int \int d^3 r d^3 r'
\rho_i({\bf r}) \rho_j({\bf r}') f_{ij}({\bf r}-{\bf r}')$, where $f_{ij}$ is
the Mayer bond between species $i$ and $j$, when all the densities are small and
the depletion potential simplifies to
\begin{equation}
\beta W({\bf r})=- \int d^3 r' (\rho_s({\bf r}')-\rho_s(\infty)) f_{bs}({\bf r}-
{\bf r}')
\end{equation}
where, in the same limit, $\rho_s({\bf r})=\rho_s(\infty) 
\exp(-\beta V_s({\bf r}))$.
For hard spheres $f_{bs}({\bf r}-{\bf r}')=-\Theta((R_s+R_b)-|{\bf r}-{\bf r}'|)$
where $\Theta$ is the Heaviside function and one can show that for a hard potential
$V_s({\bf r})$, $-\beta W({\bf r})$ is $\rho_s(\infty)$ 
times the overlap volume of the exclusion sphere of the big particle with the 
hard wall, i.e., one recovers the Asakura-Oosawa result for the depletion potential
\cite{Asakura54}.

(ii) In the limit where the two species have equal radii, i.e., $s\equiv R_s/R_b=1$, 
it is straightforward to show that $c_b^{(1)}$ reduces to the direct correlation
function of a pure ($s$) fluid and that the depletion potential is given by
$\beta W({\bf r})=-\ln(\rho_s({\bf r})/\rho_s(\infty))$, which is the exact
result \cite{Goetzelmann98b}.

Since the Rosenfeld functional is known to yield very accurate results 
\cite{Rosenfeld89} for the density profile of a pure fluid near a hard wall and
for the radial distribution function, obtained by fixing a particle at the origin
and calculating the one-body density profile, we have good reasons to expect
that the depletion potential calculated from DFT for $s\lesssim 1$ should be
rather accurate for all (fluid) densities $\rho_s(\infty)$. On the other hand for
small size ratios the accuracy of the mixture DFT is not known; we test this
by making comparison with simulations.

\begin{figure} 
\centering\epsfig{file=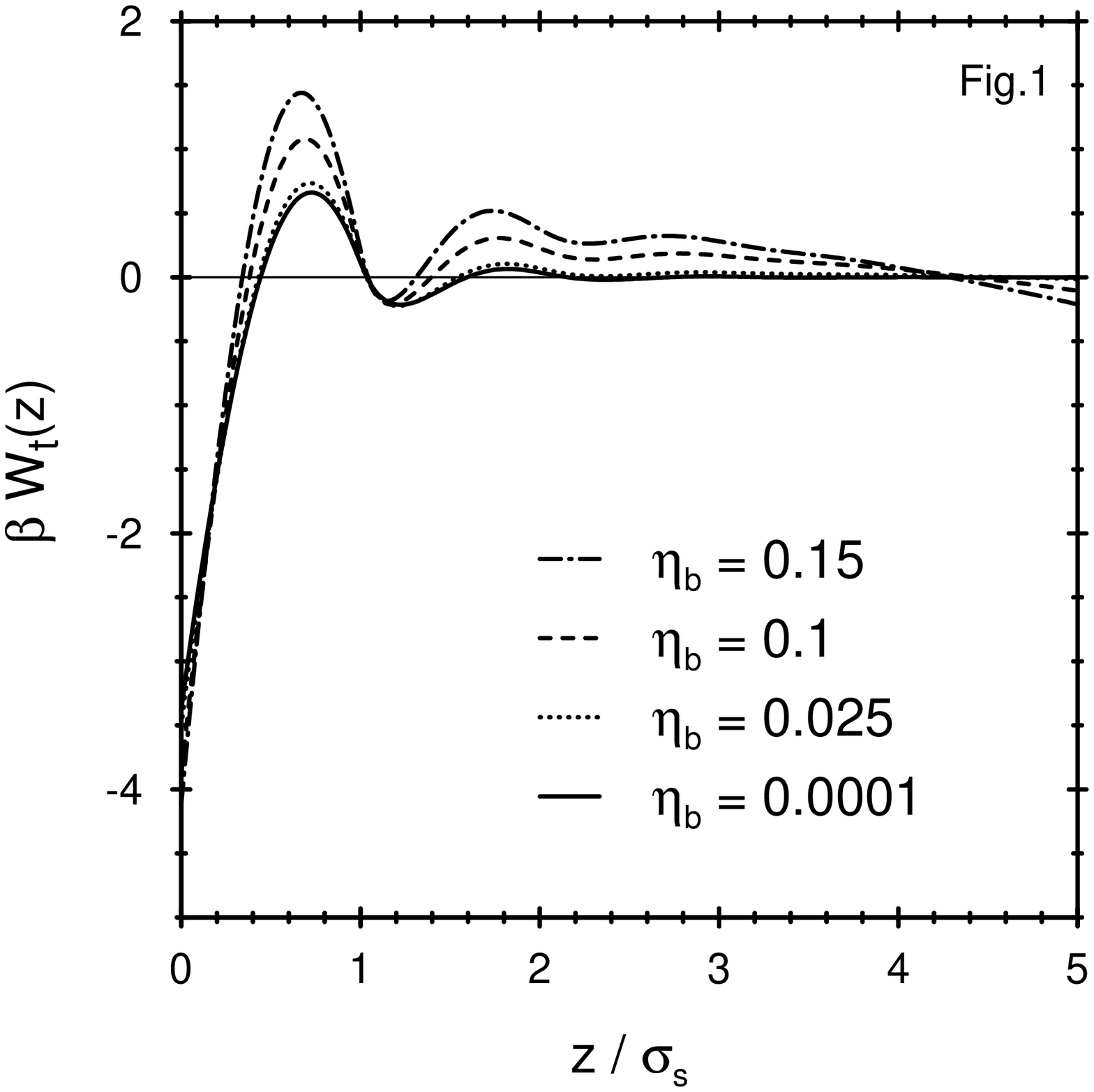,bbllx=10,bblly=60,bburx=550,bbury=580,width=13.5cm}

\vspace{0.5cm}

\caption{\label{fig1} $\beta W_t(z)$ calculated for a binary hard sphere mixture 
with size ratio $s=0.2$ and fixed small sphere packing fraction $\eta_s=0.2$ and 
for 4 different values for $\eta_b$. The separation between the surface of the big 
sphere and the hard wall is in units of the small sphere diameter $\sigma_s=2 R_s$. 
For $\eta_b=10^{-4}$, $\beta W_t(z)$ is indistinguishable from the depletion 
potential $W(z)$ obtained from the direct method.}
\end{figure}

\begin{figure}
\centering\epsfig{file=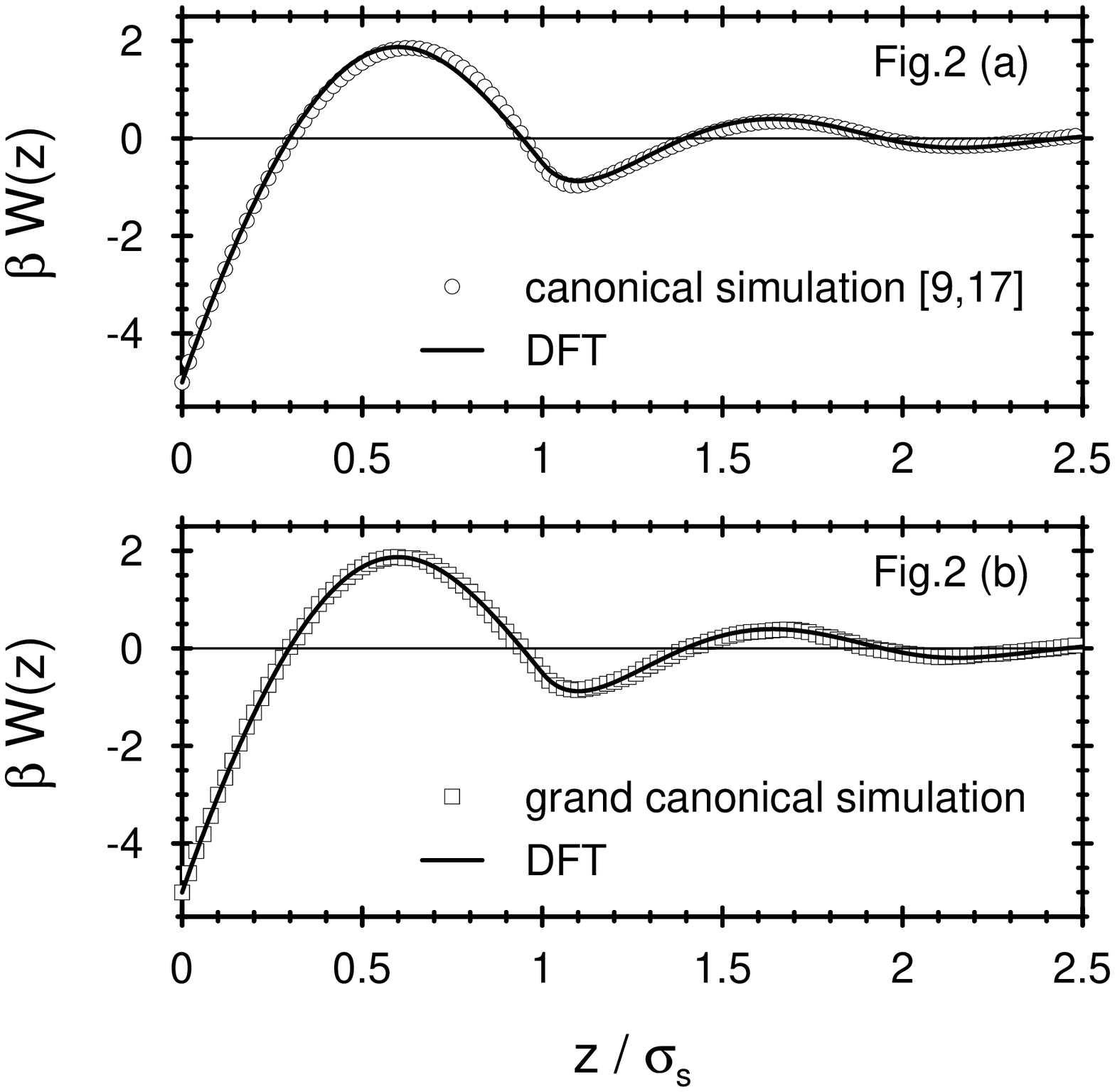,bbllx=10,bblly=60,bburx=550,bbury=580,width=13.5cm}

\vspace{0.5cm}

\caption{\label{fig2} The depletion potential $W(z)$ between a big hard sphere
and a planar hard wall for a size ratio $s=0.2$ and $\eta_s=0.3$ calculated from DFT
using the direct method (line) compared with results from 
Ref.~\protect\cite{Dickman97} (a,$\bigcirc$), shifted as described in 
Ref.~\protect\cite{comment1}, and with our present simulations (b,$\square$).}
\end{figure}

In Fig.~\ref{fig1} we show $W_t(z)$ for the binary hard sphere mixture, with
$s=0.2$, at a planar hard wall as calculated using the Rosenfeld functional for
a fixed packing fraction $\eta_s\equiv4 \pi R_s^3 \rho_s(\infty)/3 = 0.2$ of the 
small spheres and 4 different values of $\eta_b\equiv 4 \pi R_b^3 \rho_b(\infty)/3$.
$z$ is the distance between the surface of the big sphere and the wall. For 
$\eta_b=10^{-4}$, $W_t(z)$ is indistinguishable from the depletion potential $W(z)$ 
calculated using the second, direct, method. Moreover, at this concentration 
$\rho_s(z)$ is indistinguishable from the the profile of the pure fluid. As both 
functions converge fairly rapidly to their limits (the results for $\eta_b=10^{-2}$ 
differ from those for $10^{-4}$ by at most 1.6\%) we conclude that this is a 
practicable method of calculating depletion potentials via DFT. We note that for 
small values of $\eta_b$, $W_t(z)$ (and $\rho_s(z)$) exhibit decaying oscillations 
whose period is $\sim 2 R_s$, whereas for $\eta_b\gtrsim 5\times 10^{-2}$ additional,
longer period oscillations develop. The latter reflect the ordering of the big spheres
which becomes pronounced at high packings. Figure~\ref{fig2} shows the depletion
potential $W(z)$ calculated for a size ratio $s=0.2$ and $\eta_s=0.3$ alongside
the result from Ref.~\cite{Dickman97}. The latter was obtained from canonical
Monte Carlo simulations of the depletion {\em force}, as given by an integral of
the local density of small spheres in contact with the large sphere \cite{Attard89}.
Also plotted in Fig.~\ref{fig2} are our own simulation results, based on the formula
$p(z) \propto \exp(-\beta W(z))$. As direct measurement of the probability
density $p(z)$ of finding the big particle at a distance $z$ from the wall
yields poor statistics, we measured the probability ratio $p(z)/p(z+\Delta z)$ using
umbrella sampling \cite{Torrie77} in a grand canonical Monte Carlo simulation of
the small spheres, i.e., we determined $\beta (W(z+\Delta z)-W(z))$, where $\Delta z$
is a step length, and we set $W(0)$ equal to the DFT value. Our DFT results are in
excellent agreement with those of both simulations: the height of the depletion 
barrier (the maximum value of $W$ minus the contact value $W(0)$) is given almost 
exactly and the subsequent extrema are closely reproduced by the theory 
\cite{comment1}. That two independent sets of simulation results, based on totally 
different routes to the depletion potential, agree so closely is pleasing and attests 
to the accuracy with which these potentials can be calculated. We find a similar high
level of agreement between DFT and the results of Ref.~\cite{Dickman97} for
$\eta_s=0.1, 0.2$ and the smaller size ratio $s=0.1$.

The oscillatory behavior of $W(z)$ warrants further discussion. From 
the general theory of the asymptotic decay of density profiles in mixtures with
short-ranged interaction potentials it is known that the profiles of all species
near a wall (or a fixed particle) exerting a finite ranged external potential
exhibit the {\em same} characteristic decay \cite{Evans94}. Thus, for a binary mixture
near a hard wall $\rho_i(z)-\rho_i(\infty) \sim A_{wi} \exp(-a_0 z) 
\cos(a_1 z-\Theta_{wi})$, $z\to\infty$ where $i=b,s$. The amplitude $A_{wi}$
and phase $\Theta_{wi}$ depend on the particular species but $a_0$ and $a_1$
are common. In the limit $\rho_b(\infty)\to 0$ these are given by the solution
of $1-\rho_s(\infty) \hat c_{ss}(a)=0$, where $\hat c_{ss}(a)$ is the Fourier 
transform of the two-body direct correlation function of the homogeneous pure fluid
of density $\rho_s(\infty)$. $a_1$ is the real and $a_0$ the imaginary part of the
solution $a$ with the smallest imaginary part \cite{Evans94}. We have confirmed
from our numerical results that for $\eta_b=10^{-4}$ and $s=0.2$, $\rho_b(z)$ and
$\rho_s(z)$ exhibit the same value of $a_0$ and $a_1$. Moreover, for fixed
$\eta_s$, both $a_0$ and $a_1$ are, as predicted, independent of the size ratio
$s$. It follows from Eq.~(\ref{dep}) that $-\beta W(z) \sim A_{wb} \exp(-a_0 z)
\cos(a_1 z-\Theta_{wb})$, $z\to\infty$, and apart from the amplitude and phase, the
asymptotic decay of the depletion potential is also independent of $s$ 
\cite{comment2}. From a fit to the numerical data for $\eta_s=0.2$ we obtained
$a_0\approx 2.1/(2 R_s)$ and $a_1\approx 5.2/(2 R_s)$. These values are close to 
those obtained from solving $1-\rho_s(\infty) \hat c_{ss}(a)=0$ for the pure fluid 
where, consistent with the Rosenfeld DFT, $\hat c_{ss}$ is given by Percus-Yevick 
theory. Moreover, and in keeping with earlier studies of bulk pair correlation 
functions \cite{Evans94}, we find that the leading-order asymptotic result provides a
good description of the intermediate as well as the longest-ranged behavior of the 
depletion potential. We note that the period of the oscillations $2 \pi/a_1$ decreases
whereas the decay length $a_0^{-1}$ increases with increasing $\eta_s$.

The success of the comparisons between our DFT results and those of simulation lead
us to expect that the Rosenfeld functional will provide an accurate description of
depletion potentials for size ratios down to $s=0.1$ and for packing fractions
$\eta_s$ up to $0.3$. (For smaller size ratios we expect the Percus-Yevick theory
of hard sphere mixtures, upon which the functional is based, to become inaccurate
and there is no reason to expect that any approximate mixture functional would
reproduce the exact $s\to 0$ results for the depletion potential which are
obtained by making the Derjaguin approximation at the outset \cite{Goetzelmann98b}.)
This implies that our procedure can be profitably employed for some of the more
complex geometries investigated in experiments \cite{experiments}. We emphasize
that our present procedure has a crucial advantage over the brute-force application
of DFT to the calculation of depletion potentials. In the latter one would
calculate $W({\bf r})$ either from the total free energy of the inhomogeneous
fluid or from the local density of the small particles in contact with the big
one \cite{Goetzelmann98b,Attard89}. In contrast with the present scheme, both methods
require much numerical effort to minimize the free energy functional since the 
original symmetry of the density
profile $\rho_s({\bf r})$ is broken by the presence of the big particle. A
likely limitation of our procedure is that one requires a reliable free energy
functional for the (binary) {\em mixture} of big and small particles. Although
accurate functionals are available for hard {\em sphere} mixtures \cite{Rosenfeld97}
this is not the case for most other types of mixtures.

As a final remark we note that it is possible to take simulation data for
$\rho_s({\bf r})$, computed in the {\em absence} of the big test particle, and insert
these into the DFT expression Eq.~(\ref{deppot}). Although such a procedure is not 
self-consistent, in view of the high accuracy of the Rosenfeld functional it does 
provide an alternative means of calculating depletion potentials for complex
geometries such as wedges where a direct simulation of the potential or force is 
extremely difficult.

We are grateful to R. Dickman for providing his simulation results for the
depletion potential which we plot in Fig.~\ref{fig2}. We benefited from
conversations with R. Van Roij. This research was supported in part by the
EPSRC under GR/L89013.

\onecolumn

\begin{references}
\bibitem{comment1} We shifted the simulation results from Ref.~\cite{Dickman97} by a
constant  amount ($\beta W \approx 0.22$) so that the contact value matches the DFT
result. The original depletion potential did not oscillate around zero and we 
suspect this was due to the choice of cut-off $D_{max}$ used in integrating the 
force \cite{Dickman97}. After performing the shift, the simulation results oscillate
around zero. This was also the case for other choices of $\eta_s$ and for $s=0.1$.
The only exception was at $\eta_s=0.3$ and $s=0.1$, where an alternative
procedure seems necessary. This constant shift does not affect the barrier
height \cite{Dickman97}.
\end{references}
\end{document}